\documentclass[pre,citeautoscript,superscriptaddress,twocolumn,nofootinbib,floatfix]{revtex4-1}
\usepackage{xcolor}
\usepackage{graphicx}
\usepackage[utf8]{inputenc}
\usepackage[T1]{fontenc}
\usepackage{amsmath,amssymb,amsthm}
\usepackage{booktabs}
\usepackage[binary-units=true]{siunitx}
\usepackage{expl3}
\definecolor{link-color}{cmyk}{0.8 ,  0.3 ,  0. , 0}
\usepackage[colorlinks=true,%
            linkcolor   = {link-color},%
            citecolor   = {link-color},%
            urlcolor    = {link-color}]{hyperref}

\usepackage[caption=false]{subfig}

\usepackage{algpseudocode}
\usepackage{newfloat}

\DeclareFloatingEnvironment[
    fileext=loa,
    listname=List of Algorithms,
    name=ALGORITHM,
    placement=tbhp,
]{algorithm}

\makeatletter
\newcommand*{\balancecolsandclearpage}{%
  \close@column@grid
  \cleardoublepage
  \twocolumngrid
}
\makeatother

\newcommand{\V}{\mathcal{V}}
\newcommand{\Hf}{\mathcal{H}}
\renewcommand{\O}{\mathcal{O}}

\graphicspath{ {figures/} }
\begin{document}

  \author{V.A. Traag}
  \email{traag@kitlv.nl}
  \affiliation{Royal Netherlands Institute of Southeast Asian and Caribbean Studies}
  \affiliation{e-Humanities Group, Royal Netherlands Academy of Arts and
  Sciences}

  \title{Faster unfolding of communities: speeding up the Louvain algorithm.}

  \begin{abstract}

    Many complex networks exhibit a modular structure of densely connected groups of nodes.
    Usually, such a modular structure is uncovered by the optimization of some quality function.
    Although flawed, modularity remains one of the most popular quality functions.
    The Louvain algorithm was originally developed for optimizing modularity, but has been applied to a variety of methods.
    As such, speeding up the Louvain algorithm, enables the analysis of larger graphs in a shorter time for various methods.
    We here suggest to consider moving nodes to a random neighbor community, instead of the best neighbor community.
    Although incredibly simple, it reduces the theoretical runtime complexity from $\O(m)$ to $\O(n \log \langle k \rangle)$ in networks with a clear community structure.
    In benchmark networks, it speeds up the algorithm roughly $2$--$3$ times, while in some real networks it even reaches $10$ times faster runtimes.
    This improvement is due to two factors: (1) a random neighbor is likely to be in a ``good'' community; and (2) random neighbors are likely to be hubs, helping the convergence.
    Finally, the performance gain only slightly diminishes the quality, especially for modularity, thus providing a good quality-performance ratio.
    However, these gains are less pronounced, or even disappear, for some other measures such as significance or surprise. 
  
  \end{abstract}

\maketitle

\section{Introduction}

Complex networks have gained attention the past decade~\cite{boccaletti_complex_2006}. 
Especially with the rise of social media, social networks of unprecedented size became available, which contributed to the establishment of the computational social sciences~\cite{watts_twenty-first_2007,lazer_computational_2009}.
But networks are also common in disciplines such as biology~\cite{guimera_origin_2010} and neurology~\cite{betzel_multi-scale_2013}.
Many of these networks share various common characteristics. 
They often have skewed degree distributions~\cite{barabasi_scale-free_2009}, show a high clustering and a low average path length~\cite{watts_collective_1998}.
Nodes often cluster together in dense groups, usually called communities.
Nodes in a community often share other characteristics: metabolites show related functions~\cite{ravasz_hierarchical_2002} and people have a similar background~\cite{traud_social_2012}.
Revealing the community structure can thus help to understand the network~\cite{fortunato_community_2010}.

Modularity~\cite{newman_finding_2004} remains one of the most popular measures in community detection, even though it is flawed.
There have been many algorithms suggested for optimizing modularity.
The original algorithm~\cite{newman_finding_2004} created a full dendrogram and used modularity to decide on a cutting point.
It was quite slow, running in $\O(n^2 m )$, where $n$ is the number of nodes and $m$ the number of links.
Many algorithms were quickly introduced to optimize modularity, such as extremal optimization~\cite{duch_community_2005}, simulated annealing~\cite{reichardt_statistical_2006,guimera_functional_2005}, spectral methods~\cite{newman_finding_2006}, greedy methods~\cite{clauset_finding_2004}, and many other methods~\cite{fortunato_community_2010}.
One of the fastest and most effective algorithms is the Louvain algorithm~\cite{blondel_fast_2008}, believed to be running in $\O(m)$.
It has been shown to perform very well in comparative benchmark tests~\cite{lancichinetti_community_2009}.
The algorithm is largely independent of the objective function to optimize, and as such has been used for different methods~\cite{traag_narrow_2011,rosvall_mapping_2010,rosvall_multilevel_2011,ronhovde_local_2010,evans_line_2009,lancichinetti_finding_2011}

We first briefly describe the algorithm, and introduce the terminology.
We then describe our simple improvement, which we call the random neighbor Louvain, and argue why we expect it to function well.
We derive estimates of the runtime complexity, and obtain $\O(m)$ for the original Louvain algorithm, in line with earlier results, and $\O(n \log \langle k \rangle)$ for our improvement, where $\langle k \rangle$ is the average degree.
This makes it one of the fastest algorithms for community detection to optimize an objective function.
Whereas the original algorithm runs in linear time with respect to the number of edges, the random neighbor algorithm is nearly linear with respect to the number of nodes.
Finally, we show on benchmark tests and some real networks that this minor adjustment indeed leads to reductions in running time, without losing much quality.
These gains are especially visible for modularity, but less clear for other measures such as significance and surprise.

\section{Louvain algorithm}

Community detection tries to find a ``good'' partition for a certain graph.
In other words, the input is some graph $G=(V,E)$ with $n=|V|$ nodes and $m=|E|$ edges.
Each node has $k_i$ neighbors, which is called the degree, which on average is $\langle k \rangle = \frac{2m}{n}$.
The output is some partition $\V = \{V_1, V_2, \ldots, V_r\}$, where each $V_c \subseteq V$ is a set of nodes we call a community.
We work with non-overlapping nodes, such that $V_c \cap V_d = \emptyset$ for all $c \neq d$ and all nodes will have to be in a community, so that $\bigcup V_c= V$. 
Alternatively, we denote by $\sigma_i$ the community of node $i$, such that $\sigma_i = c$ if (and only if) $i \in V_c$.
Both $\sigma$ and $\V$ may be used interchangeably to refer to the partition. If the distinction is essential, we will explicitly state this.

The Louvain algorithm is suited for optimizing a single objective function that specifies some quality of a partition.
We denote such an objective function with $\Hf$, which should be maximized.
We use $\Hf(\sigma)$ and $\Hf(\V)$ to mean the same thing.
There are various choices for such an objective function, such as modularity~\cite{newman_finding_2004}, Potts models~\cite{reichardt_statistical_2006,ronhovde_local_2010,traag_narrow_2011}, significance~\cite{traag_significant_2013}, surprise~\cite{traag_detecting_2015}, infomap~\cite{rosvall_multilevel_2011} and many more.
We will not specify any of the objective functions here, nor shall we discuss their (dis)advantages, as we focus on the Louvain algorithm as a general optimization scheme.

Briefly, the Louvain algorithm works as follows. 
The algorithm initially starts out with a partition where each node is in its own community (i.e. $\sigma_i = i$), which is the initial partition.
So, initially, there are as many communities as there are nodes.
The algorithm moves around nodes from one community to another, to try to improve $\Hf(\sigma)$.
We denote by $\Delta \Hf(\sigma_i \mapsto c)$ the difference in moving node $i$ to another community $c$.
In particular, $\Delta \Hf(\sigma_i \mapsto c) = \Hf(\sigma') - \Hf(\sigma)$ where $\sigma'_j = \sigma_j$ for all $j \neq i$ and $\sigma'_i = c$, implying that if $\Delta \Hf(\sigma_i \mapsto c) > 0$, the objective function $\Hf$ is improved.
At some point, the algorithm can no longer improve $\Hf$ by moving around individual nodes, at which point it aggregates the graph, and reiterates on the aggregated graph.
We repeat this procedure as long as we can improve $\Hf(\sigma)$.
The outline of the algorithm is displayed in Algorithm~\ref{algo:louvain}.

\begin{algorithm}[t]
  \begin{algorithmic}
    \Function{Louvain}{Graph $G$}
      \State $\sigma_i \gets i$. \Comment{Initial partition}
      \State $\sigma' \gets$ \Call{MoveNodes}{$G$} \Comment{Initial move nodes}
      \While{$\Hf(\sigma') > \Hf(\sigma)$}
        \State $\sigma \gets \sigma'$
        \State $G \gets$ \Call{Aggregate}{$G, \sigma$}
        \State $\Sigma \gets $ \Call{MoveNodes}{$G$} \Comment{Move nodes}
        \State $\sigma'_i \gets \Sigma_{\sigma'_i}$ for all $i$ \Comment{Correct $\sigma'$ according to $\Sigma$}
      \EndWhile
      \State \Return $\sigma'$
    \EndFunction
    \item[]
    \Function{MoveNodes}{Graph $G$}
      \State $\sigma_i \gets i$ for $i=1,\ldots,|V(G)|$. \Comment{Initial partition}
      \State $q \gets -\infty$
      \While{$\Hf(\sigma) > q$}
        \State $q = \Hf(\sigma)$
        \For{random $v \in V(G)$}
          \State $c \gets $ \Call{SelectCommunity}{$v$} 
          \If{$\Delta \Hf(\sigma_v \mapsto c) > 0$}
            \State $\sigma_v \gets c$.
          \EndIf
        \EndFor
      \EndWhile
    \EndFunction
    \item[]
    \Function{Aggregate}{Graph $G$, Partition $\sigma$}
      \State $A \gets$ \Call{Adjacency}{$G$}
      \State $A'_{cd} \gets \sum_{ij} A_{ij} \delta(\sigma_i,c)\delta(\sigma_j,d)$
      \State \Return $A'$
    \EndFunction
  \end{algorithmic}
  \caption{Louvain method. 
  The algorithm loops over all nodes and moves nodes to alternative communities.
  When no more improvement can be made, it aggregates the graph and reiterates the procedure.
  }
  \label{algo:louvain}
\end{algorithm}

\begin{algorithm}[t]
  \begin{algorithmic}
    \item[]Select best neighbor community
    \Function{SelectCommunity}{Node $v$}
      \State $\delta \gets -\infty$.
      \State $c \gets \sigma_v$.
      \State $C \gets \{\sigma_u \mid (uv) \in E(G)\}$ \Comment{Neighbor communities.}
      \For{Community $c' \in C$}
        \If{$\Delta\Hf(\sigma_v \mapsto c') > \delta$}
          \State $\delta \gets \Delta\Hf(\sigma_v \mapsto c')$
          \State $c \gets c'$
        \EndIf
      \EndFor
      \State \Return $c$
    \EndFunction
    \item[]
    \item[]Select random neighbor community
    \Function{SelectCommunity}{Node $v$}
      \State \Return random $\sigma \in \{\sigma_u \mid (uv) \in E(G)\}$.
    \EndFunction
  \end{algorithmic}
  \caption{Select the best or a random neighbor community.} 
  \label{algo:selectcommunity}
\end{algorithm}

There are two key procedures: \textsc{MoveNodes} and \textsc{Aggregate}. 
The \textsc{MoveNodes} procedure displayed in Algorithm~\ref{algo:louvain} loops over all nodes (in random order), and considers moving them to an alternative community.
This procedure relies on \textsc{SelectCommunity} to select a (possibly) better community $c$. 
Only if the improvement $\Delta \Hf(\sigma_v \mapsto c) > 0$, we will actually move the node to community $c$.
The \textsc{Aggregate} procedure may depend on the exact quality function $\Hf$ used. 
In particular, the aggregate graph $G'$ should be constructed according to $\sigma$, such that $\Hf(G', \sigma') = \Hf(G, \sigma)$, where $\sigma'_i = i$ is the initial partition.
That is, the quality of the initial partition $\sigma'$ of the aggregated graph $G'$ should be equal to the quality of the partition $\sigma$ of the original graph $G$.
In Algorithm~\ref{algo:louvain} a version is displayed which is suited for modularity.
Other methods may require additional variables to be used when aggregating the graph (e.g.~\cite{traag_narrow_2011}).

The only procedure that remains to be specified is \textsc{SelectCommunity}. 
In the original Louvain algorithm, this procedure commonly considers all possible neighboring communities, and then greedily selects the best community.
It is summarized in Algorithm~\ref{algo:selectcommunity}.

We created a new flexible and fast implementation of the Louvain algorithm in \texttt{C++} for use in \texttt{python} using \texttt{igraph}.
The implementation of the algorithm itself is quite detached from the objective function to optimize.
In particular, all that is required to implement a new objective function is the difference when moving a node $\Delta \Hf$ and the quality function $\Hf$ itself (although the latter is not strictly necessary).
This implementation is available open source from~\texttt{GitHub}\footnote{\url{https://github.com/vtraag/louvain-igraph}} and~\texttt{PyPi}\footnote{\url{https://pypi.python.org/pypi/louvain}}.

\section{Improvement}

\begin{figure*}[t]
  \begin{center}
    \includegraphics{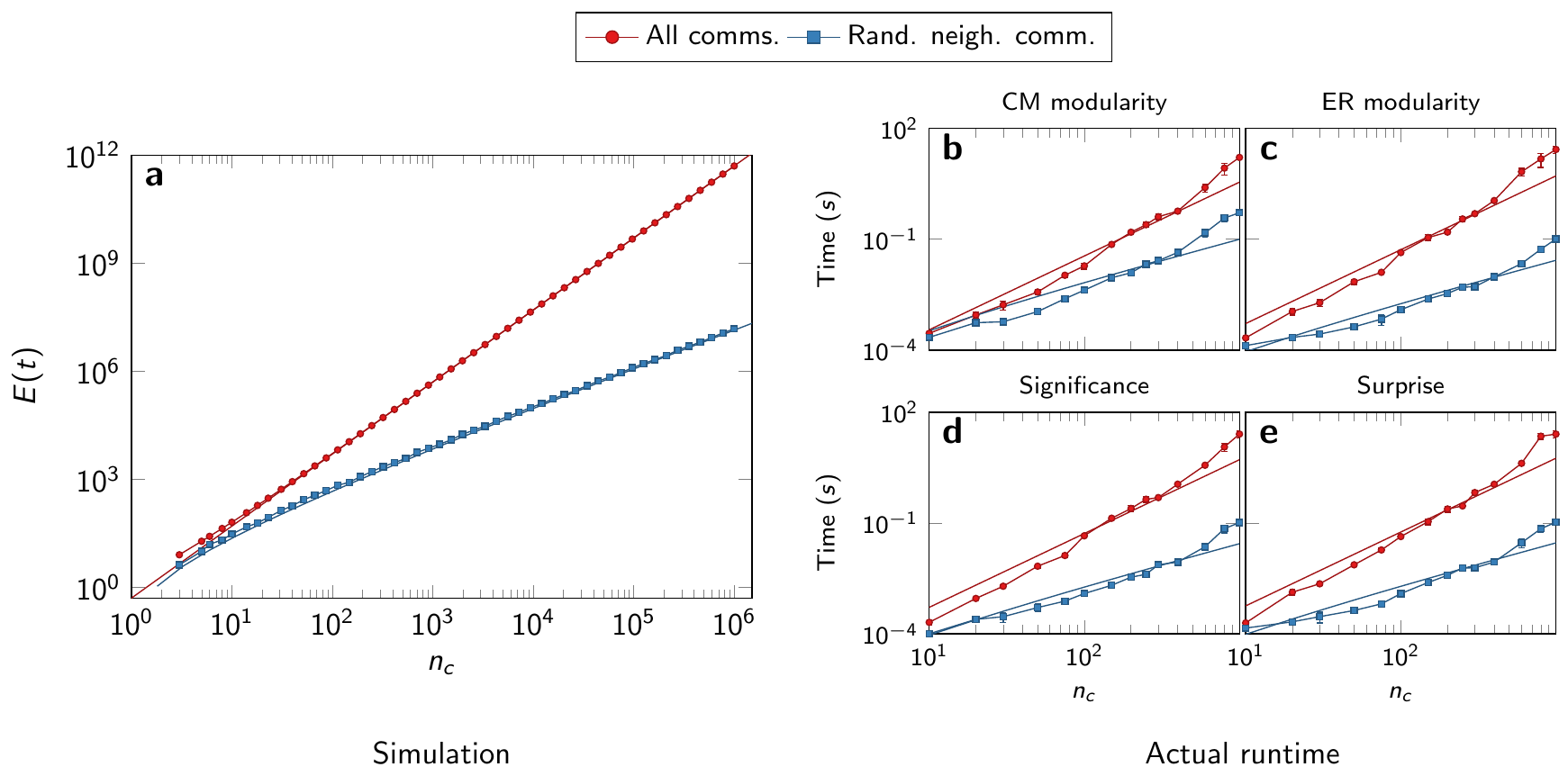}
  \end{center}
  \caption{\textbf{Clique.} 
    The original Louvain algorithm considers all communities, which leads to $E(t) = \O(n_c^2)$ operations for putting all $n_c$ nodes of a clique in a single community.
    The improvement considers only random neighbors, which takes only $E(t) = \O(n_c \log n_c)$ operations to identify the whole clique as a community. 
    In (a) we show the number of operations in a simulation, with the markers indicating the simulated number of operations, and the solid lines the analytically derived estimates.
    In (b)--(e) we show the actual time used when optimizing the indicated quality functions for a clique for the different objective functions.
    The solid lines in (b)--(e) denote best fits to $n_c^2$ and $n_c \log n_c$ in log-space.
  }
  \label{fig:clique}
\end{figure*}

Not surprisingly, the Louvain algorithm generally spends most of its time contemplating alternative communities.
While profiling our implementation, we found that it spends roughly $95\%$ of the time calculating the difference $\Delta \Hf(\sigma_v \mapsto c)$ in Algorithm~\ref{algo:selectcommunity}.
Much of this time is spent moving around nodes for the first time.
With an initial partition where each node is in its own community, almost any neighboring community would be an improvement.
Moreover, when the algorithm has progressed a bit, many neighbors likely belong to the same community.
We therefore suggest that instead of considering all neighboring communities, we simply select a random neighbor, and consider that community (as stated in Algorithm~\ref{algo:selectcommunity}), which we call the random neighbor Louvain.
Notice that the selection of a random neighbor makes the greedy Louvain algorithm less greedy and thus more explorative.
Indeed, when also accepting moves with some probability depending on the improvement (possibly also accepting degrading moves), the algorithm comes close to resemble simulated annealing~\cite{reichardt_statistical_2006,guimera_functional_2005}.
However, simulated annealing is rather slow for community detection~\cite{lancichinetti_community_2009}, so we don't explore that direction further, since we are interested in speeding up the algorithm.

There are several advantages to the selection of a random neighbor.
First of all, it is likely to choose a relatively ``good'' community.
In general, a node should be in a community to which relatively many of its neighbors belong as well (although this of course depends on the exact quality function).
By selecting a community from among its neighbors, there is a good chance that a relatively good community is picked.
In particular, if node $i$ has $k_i(c)$ neighbors in community $c$, the probability that community $c$ will be considered for moving is $k_i(c)/k_i$.
The probability for selecting a community is thus proportional to the number of neighbors in that community.
Bad communities (with relatively few neighbors) are less frequently sampled, so that the algorithm focuses more on the promising communities (those with relatively many neighbors).

Moreover, when considering the initial partition of each node in its own community, almost any move would improve the quality function $\Hf$.
The difference between alternative communities in this early stage is likely to be marginal.
Any move that puts two nodes in the same community is probably better than a node in its own community. 
Such moves quickly reduce the number of communities from roughly $n$ to $n/2$.
But instead of considering every neighboring community as in the original Louvain algorithm, which takes roughly $\O(\langle k \rangle)$, our random neighbor Louvain algorithm only considers a single random neighbor, which takes constant time $\O(1)$.
So, for the first few iterations, Louvain runs in $\O(n \langle k \rangle) = \O(m)$, whereas selecting a random neighbor runs in $\O(n)$.

\begin{figure}[t]
  \begin{center}
    \includegraphics{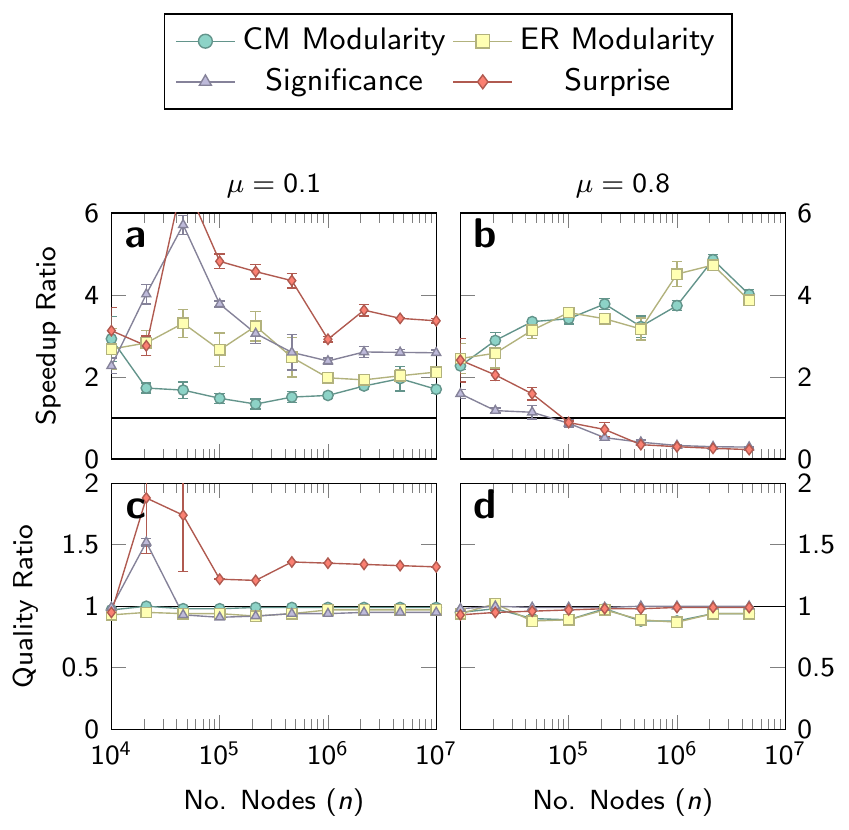}
  \end{center}
  \caption{\textbf{Network size.}
    We here show the ratio of the time and of the quality (i.e. $\Hf$) of the uncovered partitions by the original Louvain algorithm and the random neighbor Louvain.
    The random neighbor Louvain algorithm is $2$--$3$ times faster than the original Louvain algorithm and at some points event faster for clear communities in (a) when using $\mu = 0.1$.
    However, for less clear communities at $\mu = 0.8$ as displayed in (b), the optimization of significance and surprise is not faster by using the random neighbor Louvain.
    The random neighbor Louvain uncovers almost the same quality as the original version for a large part, as shown in (c) and (d).
    However, especially for surprise, the quality is adversely affected by the random neighbor Louvain for $\mu=0.1$, shown in (c).
    The results are based on benchmark graph with communities of size $n_c = 1\,000$ and an average degree of $\langle k \rangle = 15$.
  }
  \label{fig:performance}
\end{figure}

\begin{figure}[t]
  \begin{center}
    \includegraphics{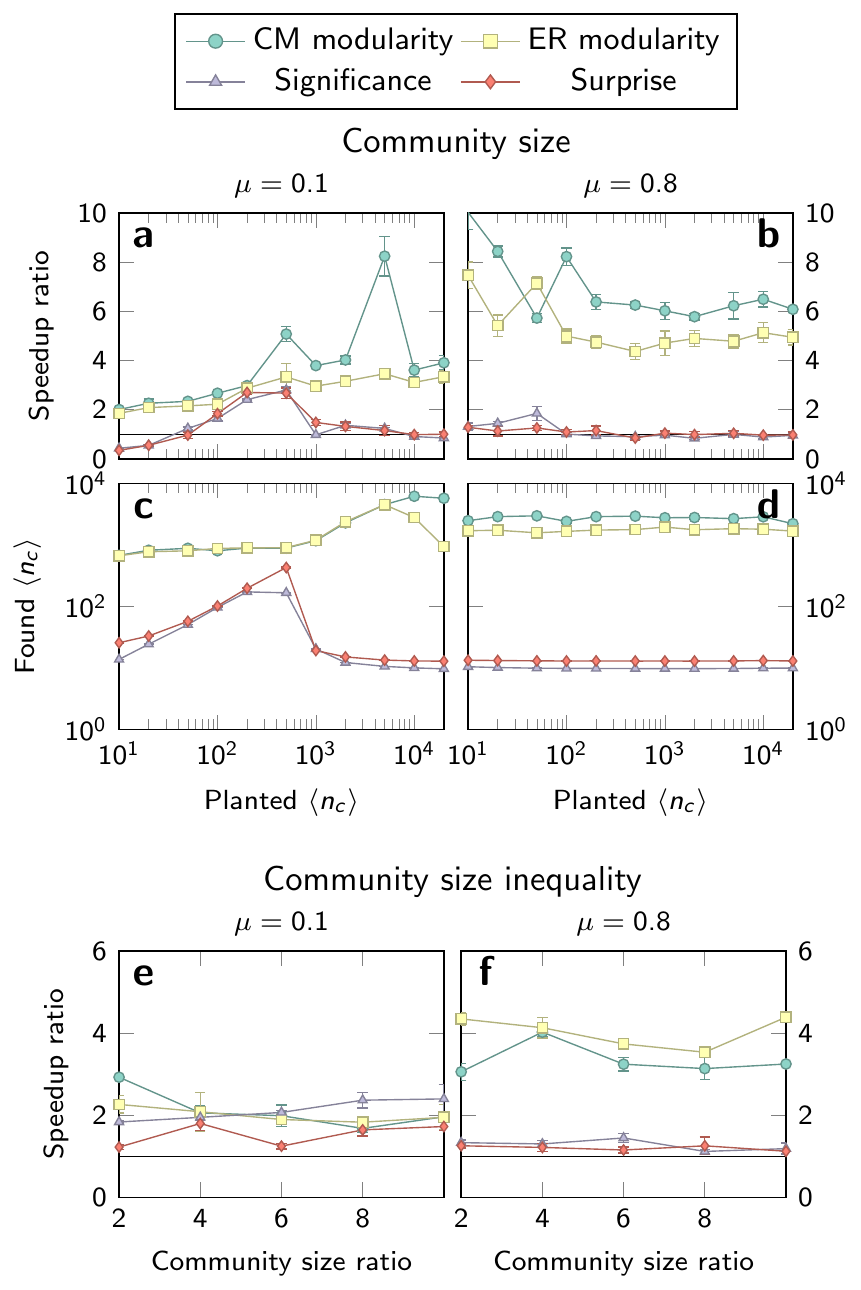}
  \end{center}
  \caption{\textbf{Effect of community size.}
    Results in (a) show that the speedup ratio increases with the community size for $\mu = 0.1$.
    Surprise and significance find smaller substructures within large communities as seen in (c).
    This is also when the improvement starts to deteriorate.
    The results for large communities in (a) and (c) are very similar to the situation when $\mu = 0.8$ in (b) and (d), which resembles a random graph more closely.
    The speedup ratio in (b) corresponds to this: the speedup is rather large for modularity, while it is much lower for surprise and significance.
    In (e) and (f) we show that heterogeneity in the community sizes nearly do not impact the speedup ratio.
    In that case we generate LFR benchmark graphs with smallest community size $n_c = 10$ and the maximum community size varies from $2$ to $10$ times as large.
    We use $n = 10^5$ and $\langle k \rangle = 10$ for both benchmarks.
  }
  \label{fig:performance_comm_size}
\end{figure}

Notice there is a big difference between (1) selecting a random neighbor and then its community and (2) selecting a random community from among the neighboring communities.
The first method selects a community proportional to the number of neighbors that are in that community, while the second method selects a community uniformly from the set of neighboring communities.
Consider for example a node that is connected to two communities, and has $k_i - 1$ neighbors in the first community and only $1$ in the other community.
When selecting a community of a random neighbor, the probability the good community is considered is $1 - \frac{1}{k_i}$, while the probability is only $\frac{1}{2}$ when selecting a random community.

Secondly, random selection of a neighbor increases the likelihood of quick convergence.
The probability that node $i$ is selected as a random neighbor is roughly $k_i/2m$, resembling preferential attachment~\cite{barabasi_emergence_1999} in a certain sense.
Hubs are thus more likely to be chosen as a candidate community.
Since, hubs connect many vertices, there is a considerable probability that two nodes consider the same hub.
If these two (or more) nodes (and the hub) should in fact belong to the same community, chances are high both nodes and the hub quickly end up in the same community.

As an illustration of this advantage, consider a hubs-and-spokes structure, with one central hub and only neighboring spokes that are connected to each other (and always to the hub).
So, any spoke node $i$ is connected to nodes $i - 1$ and $i + 1$ and to the central hub, node $n$.
Consider for simplicity that the nodes are considered in order and that every move will be advantageous.
The probability that the first node will move to community $n$ is $p_1 = \frac{1}{3}$.
For the second node, he will move to community $n$ if he chooses node $n$ immediately (which happens with probability $\frac{1}{3}$), or if he chooses node $1$, and node $1$ moved to community $n$, so that $p_2 = \frac{1}{3} + p_1\frac{1}{3}$.
Similarly, for the other nodes $p_i = \frac{1}{3} + p_{i-1} \frac{1}{3} = \sum_{j=1}^i \left(\frac{1}{3}\right)^j$ which goes to $\frac{1}{2}$ for $n \to \infty$.
This is higher than when just considering a random neighbor community.
In that case, the probability the first node will move to community $n$ is still $\frac{1}{3}$.
But for the second node, if node $1$ moved to community $n$, only two communities are left: $n$ and $3$.
In that case, community $n$ is chosen with probability $\frac{1}{2}$.
If node $1$ didn't move to community $n$, then node $2$ will move to community $n$ with probability $\frac{1}{3}$.
In general, node $i$ moves to community $n$ with probability $p_i = p_{i-1} \frac{1}{2} + (1 - p_{i-1}) \frac{1}{3} = p_{i-1}\frac{1}{6} + \frac{1}{3}$.
Working out the recurrence, we obtain that $p_i = \frac{1}{3} \sum_{j=0}^{i-1} \left(\frac{1}{6}\right)^j$, which tends to $\frac{2}{5}$.
Selecting a community of a random neighbor thus works better than selecting a random community from among the neighbors.
Selecting a community of a random node is even worse.
In that case, the probability is $p_i = \frac{1}{n}(1 + \frac{1}{n})^{i-1}$ which tends to $0$ for $n \to \infty$.
In short, selecting the community of a random neighbor is likely to choose a new community that will also be chosen by other nodes.

In summary, selecting a random neighbor should work well because of two reasons.
First, it tends to focus on communities that are ``good''.
Secondly, it should help in convergence because of higher likelihood of selecting hubs.
In particular, the evaluation of $\textsc{SelectCommunity}$ in the random neighbor Louvain takes a constant time $\O(1)$ whereas evaluating all communities takes about $O(\langle k \rangle)$.
However, one essential question is whether $\textsc{SelectCommunity}$ will not be too frequently evaluated in the random neighbor Louvain to counter this benefit. 

\begin{figure*}[t]
  \begin{center}
    \includegraphics{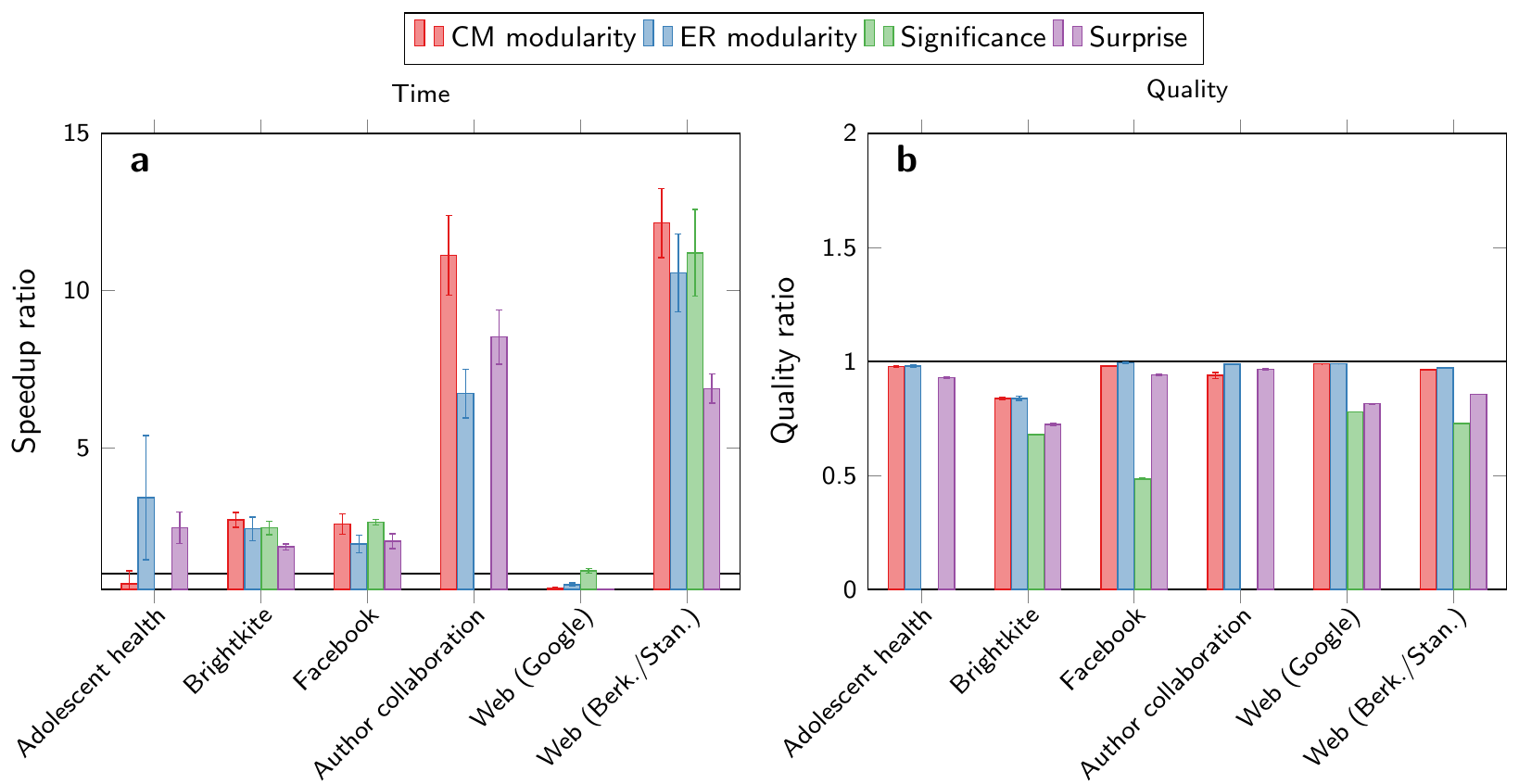}
  \end{center}
  \caption{\textbf{Empirical network results.} 
    The random neighbor Louvain usually speeds up the optimization of the objective function for most empirical networks. 
    For the hyperlink network from Google it does not work for any method, while the adolescent health dataset poses problems for optimizing CM modularity.
    The quality remains relatively similar compared to the original, especially for modularity as shown in (b).
    For significance and surprise the difference are more pronounced.
  }
  \label{fig:real_networks}
\end{figure*}

\begin{table}
  \sisetup{
    table-number-alignment=center,
    group-minimum-digits=3} 
  \begin{tabular}{
      l
      S[table-format=7.1]
      S[table-format=8.1]
      S[table-format=4.2]}
    \toprule 
    Network               & {$n$}    & {$m$}     & {$\langle k \rangle$} \\
    \midrule
    Health                & 2539   & 12969   & 10.22 \\
    Brightkite            & 58228  & 214078  & 7.35 \\
    Facebook              & 63731  & 817035  & 25.64 \\
    Author Collaboration  & 22908  & 2673133 & 233.38 \\
    Web (Google)          & 875713 & 5105039 & 11.66 \\
    Web (Berk./Stan.)     & 685230 & 7600595 & 22.18 \\
    \bottomrule
  \end{tabular}
  \caption{\textbf{Empirical network overview.}}
  \label{tab:real_networks}
\end{table}

To study this question, let us consider a ring of $r$ cliques of $n_c$ nodes each.
The cliques (which are complete subgraphs containing $\binom{n_c}{2}$ links) are connected to another clique only by a single link in a circular fashion (i.e. clique $i$ is connected only to clique $i-1$ and $i+1$).
Most methods tend to find the cliques (or sets of multiple cliques due to the resolution limit~\cite{fortunato_resolution_2007,traag_narrow_2011}).
Indeed, it is one of the best possible community structures: we cannot add any more internal edges, nor can we delete any external edges without disconnecting the graph.
However, for the runtime complexity, the external edges will play only a marginal role.
We may therefore simply assume we will work with $r$ disconnected cliques of size $n_c$.
Although the actual runtime will deviate from this, it should provide a reasonable runtime for relatively ``clear'' communities, and as such provide a lower bound for more difficult communities.

The core question is thus how quickly both the original and the random neighbor Louvain run on cliques.
We will assume the clique should become a single community, which is likely to be the case for most methods.
Additionally, we assume $\Delta \Hf > 0$ only if a node is moved to a larger community, which is likely to be the case for most methods as nodes in a clique have more links to larger communities.
The complexity of the original Louvain implementation is simple to evaluate in this case.
The first node will be moved to one of its neighbors, an operation that costs $n_c$ evaluations.
The second node has only $n_c-1$ evaluations to make, since the community of the first node disappeared.
If we continue in this fashion, the total number of evaluations $t$ is then 
$\sum_{i = 1}^{n_c} n_c - i + 1 = \frac{n_c(n_c + 1)}{2} = \O(n_c^2)$.
The analysis of the expected runtime of the random neighbor Louvain is more difficult (see Appendix~\ref{sec:clique} for more details).
However, we can provide a lower bound that serves as a rough estimate.
Let us again denote by $t$ the total number of operations before the whole clique is identified as a single community.
We divide this in different phases of the algorithm, where each phase $i$ runs from the time where there are $n_c - i + 1$ communities, until there are $n_c - i$ communities.
In phase $1$ we thus start out with $n_c$ communities, and in the next phase there are only $n_c - 1$ communities.
If we denote by $t_i$ the number of operation in phase $i$, then by linearity $E(t) = E(\sum_t t_i) = \sum_t E(t_i)$.
Notice that we will only leave phase $i$ whenever a community of size $1$ disappears.
The probability that a community of $1$ disappears is $\frac{n_c - 1}{n_c}$, since it will join any other community (except itself).
There are at most $i$ communities of size $1$ in phase $i$, so that the probability a community of size $1$ is selected is bounded above by $\frac{n_c - i + 1}{n_c}$.
In fact, such a state is also relatively likely, as the community size distribution tends to become more skewed than a more uniform distribution due to the preferential attachment on the basis of the community sizes.
The number of expected operations in phase $i$ is then bounded below by $\frac{n_c}{n_c - i + 1}$, and the expected operations in total is bounded below by 
\begin{align}
  E(t) &\geq n_c \sum_{i = 1}^{n_c} \frac{1}{n_c - i + 1} \\
       &= n_c \sum_{i = 1}^{n_c} \frac{1}{i} = \O(n_c \log n_c)
\end{align}
However, this lower bound gives in fact a very accurate estimate of the expected running time, as seen in Fig.~\ref{fig:clique}.
Whereas the original Louvain algorithm runs in $\O(n_c^2)$, the random neighbor version only uses $\O(n_c \log n_c)$ to put all nodes of a clique in a single community.
We used an explicit simulation of this process to validate our theoretical analysis.
Running the actual algorithms on cliques yields similar results (Fig.~\ref{fig:clique}).

To get a rough idea of the overall running time, let us translate these results back to the ring of cliques.
In that case, we have $r$ cliques of $n_c$ nodes.
The runtime for the original Louvain method is $\O(n_c^2)$ for each clique, so that the total runtime is about $\O(rn_c^2)$.
One factor of $n_c^2$ comes from running over $n_c$ nodes, while the other factor comes from running over $\langle k \rangle \approx n_c$ neighbors.
Since $rn_c = n$, and $n \langle k \rangle = m$, we thus obtain an overall running time of Louvain of about $\O(rn_c^2) = \O(n \langle k \rangle) = \O(m)$, similar to earlier estimates~\cite{blondel_fast_2008,fortunato_community_2010}.
Following the same idea, we obtain an estimate of roughly $\O(n \log \langle k \rangle)$ for the runtime of the random neighbor Louvain algorithm.
So, whereas the original algorithm runs in roughly linear time with respect to the number of edges, the random neighbor algorithm runs in nearly linear time with respect to the number of nodes.
Empirical networks are usually rather sparse, so that the difference between $\langle k \rangle$ and $\log \langle k \rangle$ is usually not that large.
Still, it is quite surprising to find such an improvement for such a minor adjustment.

\section{Experimental Results}

We use benchmark networks and real networks to show that the random neighbor improvement also reduces the runtime in practice.
These benchmark networks contain a planted partition, which we then try to uncover using both the original and the random neighbor Louvain algorithm.
An essential role is played by the probability that a link falls outside of the planted community $\mu$.
For low $\mu$ it is thus quite easy to identify communities, while for high $\mu$ it becomes increasingly more difficult.
We report results using the speedup ratio calculated as $R_\text{speed} = \frac{T_\text{orig}}{T_\text{rn}}$, where $T_\text{rn}$ is the runtime of the random neighbor variant and $T_\text{orig}$ the runtime of the original Louvain method. 
The runtime is calculated in used CPU time, not elapsed real time.
We also report the quality ratio, which is calculated as $R_\text{qual} = \frac{\Hf_\text{rn}}{\Hf_\text{orig}}$ where $\Hf_\text{rn}$ refers to the quality of the partition uncovered using the random neighbor improvement and $\Hf_\text{rn}$ to the quality using the original algorithm.
In this way, if $R_\text{speed} > 1$ the random neighbor improves upon the original and similarly $R_\text{qual} > 1$ if the random neighbor is an improvement.
Throughout all plots, error bars indicate standard errors of the mean.
The Louvain algorithm can be applied to many different methods, and we here show results for (1) modularity using a configuration null model~\cite{newman_finding_2004} (CM modularity); (2) modularity using an Erdös-Rényi null model~\cite{reichardt_statistical_2006} (ER modularity); (3) significance~\cite{traag_significant_2013}; and (4) surprise~\cite{aldecoa_surprise_2013}.

We first test the impact of the network size as a whole.
We construct benchmark networks ranging from $n=10^4$ to $n=10^7$ nodes, with equally sized communities of $1\,000$ nodes, with a Poissonian degree distribution.
The speed and quality of the original Louvain algorithm and the random neighbor Louvain algorithm for all four methods is reported in Fig.~\ref{fig:performance}.
For all these methods, the random neighbor Louvain speeds up the algorithm roughly $2$--$3$ times.
At the same time, the quality of the partitions found remains nearly the same.

However, surprise and significance seem to perform worse than modularity.
The speedup is rather limited for higher $\mu$ (or becomes even slower than the original), in which case communities are more difficult to detect.
Surprise and significance tend to find relatively smaller communities than modularity~\cite{traag_detecting_2015}, suggesting that the performance gain of using the random neighbor Louvain is especially pertinent when making a relatively coarse partition.
Revisiting the argument of the ring of cliques makes clear that the runtime does not necessarily scale with the degree, but rather, with the clique size, which we may approximate as the community size.
Indeed, the runtime for merging all the $n_c$ nodes in a single community together, should take $\O(n_c^2)$ originally and $\O(n_c \log n_c)$ in the random neighbor Louvain, as previously argued.
However, if there are no clear communities present in the network, the running time will not depend on the degree as much, but rather on the sizes of the communities found.
Hence, the running time should then roughly scale as $\O(n n_c)$ for the original implementation and as $\O(n \log n_c)$ for the random neighbor Louvain.
Since surprise and significance find smaller communities than modularity (unless the communities are clearly defined), the speedup will be rather limited, whereas it will be larger generally for modularity.

We test this by generating benchmark networks with $n=10^5$ nodes, $\langle k \rangle = 10$ and varying community sizes from $10$ to $20\,000$.
Results are displayed in Fig.~\ref{fig:performance_comm_size}.
Indeed, for larger communities, surprise and significance have difficulties discerning such large communities, and it tends to find substructure within these large communities.
Notice that modularity also merges smaller communities (thereby uncovering artificially larger communities), part of the problem of the resolution limit~\cite{fortunato_resolution_2007}.
This is exactly also the point at which the speedup for surprise and significance goes down.
Moreover, when the community structure is not clear, there is no effect of community size at all.
Indeed, in that case, surprise tends to find small communities, and modularity tends to find large communities.
The speedup follows this pattern: surprise and significance show very small speedups, while modularity shows larger speedups.

However, modularity also prefers rather balanced communities~\cite{Lancichinetti2011}, so that perhaps modularity performs rather well because of the similarity in community sizes.
We therefore also consider the impact of more heterogeneity by constructing LFR benchmark networks~\cite{lancichinetti_benchmark_2008}.
In these benchmark graphs the community sizes and the degree both follow powerlaw distributions with exponents $1$ and $2$ respectively.
The maximum degree was set at $2.5 \langle k \rangle$, while the minimum community size was set at $\langle k \rangle$ for $\langle k \rangle = 10$. 
We varied the maximum community size from $2 \langle k \rangle$ to $10 \langle k \rangle$.
These results are displayed in Fig.~\ref{fig:performance_comm_size}, from which we can see that the heterogeneity in community sizes does not affect the results.

We also tested the random neighbor Louvain on six empirical networks of varying sizes.
These networks were retrieved from the Koblenz Network Collection\footnote{\url{http://konect.uni-koblenz.de/}}.
We include 
(1) the \href{http://konect.uni-koblenz.de/networks/moreno_health}{adolescent health dataset}, a school network collected for health research~\cite{Moody2001};
(2) \href{http://konect.uni-koblenz.de/networks/loc-brightkite_edges}{Brightkite}, a social network site~\cite{Cho2011};
(3) a \href{http://konect.uni-koblenz.de/networks/facebook-wosn-links}{Facebook} friendship network~\cite{Viswanath2009};
(4) an \href{http://konect.uni-koblenz.de/networks/ca-cit-HepPh}{author collaboration network} from the High Energy topic on arXiv~\cite{JureLeskovecJonKleinberg2006};
(5) a \href{http://konect.uni-koblenz.de/networks/web-Google}{web hyperlink network} released by Google~\cite{Leskovec2008}; and 
(6) the \href{http://konect.uni-koblenz.de/networks/web-BerkStan}{complete web hyperlink network} from the universities of Berkeley and Stanford~\cite{Leskovec2008}.
An overview of the size of the networks is provided in Table~\ref{tab:real_networks}, and the results are displayed in Fig.~\ref{fig:real_networks}.
The random neighbor Louvain is clearly faster for most networks and methods, reaching even speedup ratios of over 10 for the hyperlink web network from Berkeley and Stanford.
For the web network released by Google the improvement is not faster however.
The quality remains relatively similar for most networks, especially for modularity, whereas the quality differs more for surprise and significance.

\begin{figure}[t]
  \begin{center}
    \includegraphics{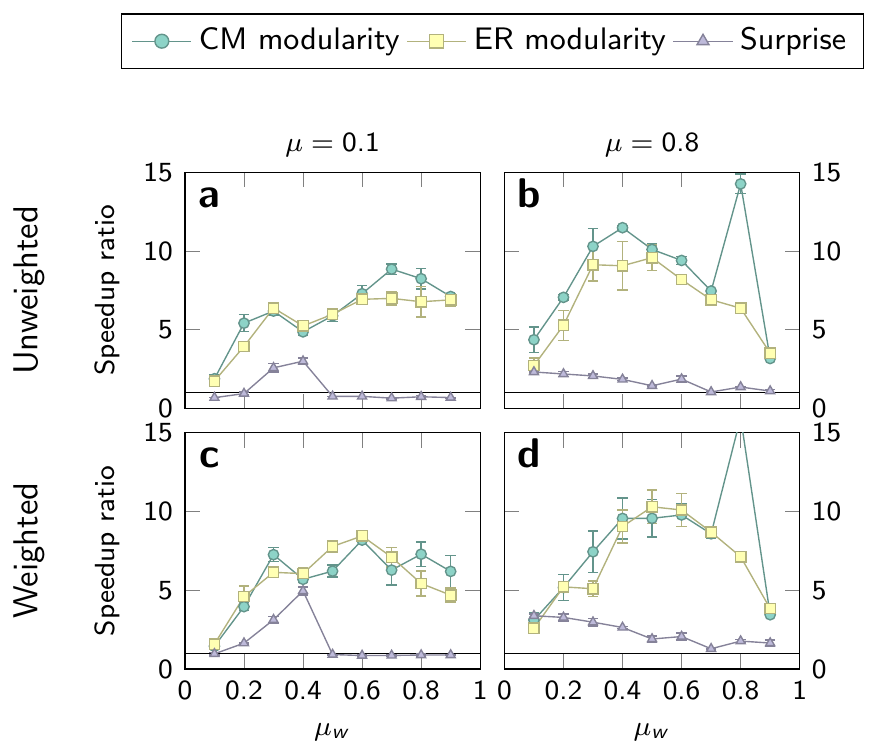}
  \end{center}
  \caption{\textbf{Performance weighted neighbor sampling.} 
    Instead of sampling a neighbor randomly, it is also possible to sample neighbors proportional to the weight.
    We here test the performance of the unweighted neighbor sampling in (a)--(b) and the weighted neighbor sampling in (c)--(d).
    We generate weighted LFR benchmark networks, where the strength of the nodes follows the degree $s_i = k_i^\beta$ with $\beta = 1.5$ with $\langle k \rangle = 10$ and $n=10^5$.
    The results for the unweighted neighbor sampling in (a) and (b) are very similar to the results for the weighted neighbor sampling in (c) and (d).
    Taking into account the weight hence does not improve the random neighbor sampling much.
  }
  \label{fig:performance_weighted}
\end{figure}

Notice that significance is not defined for weighted networks, such that significance is not run on those networks (health and author collaboration).
But weighted networks raise an interesting point: is it possible to make use of the weight to improve the speed even more?
A natural possibility is to sample neighbors proportional to the weight.
Neighbors in the same community are often connected with a higher weight, part of the famous strength of weak ties~\cite{Granovetter1973,Onnela2007a}.
Sampling proportional to the weight should thus increase the chances of drawing a ``good'' community.
However, this depends on the extent to which this correlation between weight and community holds.
The aggregated graph is weighted also, allowing the possibility of weighted sampling as well. 
On the other hand, only little time is spent in the aggregated iterations, making the benefit relatively small.
Weighted sampling in constant time requires preprocessing, which takes an additional $\O(m)$ memory and $\O(m)$ time.
The question is thus whether these costs do not offset the possible benefits.

We use weighted benchmark networks~\cite{Lancichinetti2009} to test whether weighted sampling speeds up the algorithm even further.
These benchmark networks introduce an additional mixing parameter for the weight $\mu_w$.
Whereas the topological mixing parameter $\mu$ controls the probability of an edge outside of the community, the weight is distributed such that on average a proportion of about $\mu_w$ lies outside of the community.
The strength of the nodes follows the degree $s_i = k_i^\beta$ with $\beta = 1.5$ with $\langle k \rangle = 10$ and $n=10^5$.
The external weight $\mu_w s_i$ is spread over $\mu k_i$ external links, thereby leading to an average external weight of $\frac{\mu_w s_i}{\mu k_i} = \frac{\mu_w}{\mu} k_i^{\beta - 1}$.
If $\mu_w > \mu$ the external weight is higher than the internal weight, making it difficult to detect communities correctly.
Intuitively, we would thus expect to see an improvement in the random neighbor selection whenever $\mu_w < \mu$, as in that case, the weight correlates with the planted partition.
The results for both the unweighted and the weighted random neighbor sampling is displayed in Fig.~\ref{fig:performance_weighted}.
Although the weighted random neighbor sampling sometimes improves on the unweighted variant, overall the performance is comparable. 
The results on the unweighted benchmark networks and the empirical networks are also very comparable (not shown).

\section{Conclusion}

Many networks seem to contain some community structure.
Finding such communities is important across many different disciplines.
One of the most used algorithms to optimize some quality function is the Louvain algorithm.
We here showed how a remarkably simple adjustment leads to a clear improvement in the runtime complexity.
We argue that the approximate runtime of the original Louvain algorithm should be roughly $\O(m)$, while the improvement reduces the runtime to $\O(n \log \langle k \rangle)$ in a clear community structure.
So, whereas the original algorithm is linear in the number of edges, the random neighbor algorithm is nearly linear in the number of nodes.

We have tested the random neighbor algorithm extensively.
The improvement is quite consistent across various settings and sizes.
The runtime complexity was reduced, speeding up the algorithm roughly $2$--$3$ times, especially when concentrating on the coarser partitions found by modularity. 
Nonetheless, some methods, such as surprise and significance, are more sensitive to sampling a random neighbor.
This seems to be mostly due to the community size in the uncovered partition.
Whereas modularity prefers rather coarse partitions, both significance and surprise prefer more refined partitions, leading to much smaller communities.
More refined partitions offer fewer opportunities for improving the runtime, so that sampling a random neighbor provides little improvement. 

The idea could also be applied in different settings.
For example, the label propagation method is also a very fast algorithm~\cite{raghavan_near_2007}, but it doesn't consider any objective function.
It simply puts a node in the most frequent neighboring community.
But instead of considering every neighbor, it can simply choose a random neighbor, similar to the improvement here.
We may thus expect a similar improvement in label propagation as for the Louvain algorithm.
Similar improvements may be considered in other algorithms.
The core of the idea is that a random neighbor is likely to be in a ``good'' community, which presumably also holds for other algorithms.

\begin{acknowledgments}
This research is funded by the Royal Netherlands Academy of Arts and Sciences (KNAW) through its eHumanities project~\footnote{\url{http://www.ehumanities.nl/computational-humanities/elite-network-shifts/}}.
\end{acknowledgments}

\bibliography{bibliography_formatted}

\begin{thebibliography}{41}%
\makeatletter
\providecommand \@ifxundefined [1]{%
 \@ifx{#1\undefined}
}%
\providecommand \@ifnum [1]{%
 \ifnum #1\expandafter \@firstoftwo
 \else \expandafter \@secondoftwo
 \fi
}%
\providecommand \@ifx [1]{%
 \ifx #1\expandafter \@firstoftwo
 \else \expandafter \@secondoftwo
 \fi
}%
\providecommand \natexlab [1]{#1}%
\providecommand \enquote  [1]{``#1''}%
\providecommand \bibnamefont  [1]{#1}%
\providecommand \bibfnamefont [1]{#1}%
\providecommand \citenamefont [1]{#1}%
\providecommand \href@noop [0]{\@secondoftwo}%
\providecommand \href [0]{\begingroup \@sanitize@url \@href}%
\providecommand \@href[1]{\@@startlink{#1}\@@href}%
\providecommand \@@href[1]{\endgroup#1\@@endlink}%
\providecommand \@sanitize@url [0]{\catcode `\\12\catcode `\$12\catcode
  `\&12\catcode `\#12\catcode `\^12\catcode `\_12\catcode `\%12\relax}%
\providecommand \@@startlink[1]{}%
\providecommand \@@endlink[0]{}%
\providecommand \url  [0]{\begingroup\@sanitize@url \@url }%
\providecommand \@url [1]{\endgroup\@href {#1}{\urlprefix }}%
\providecommand \urlprefix  [0]{URL }%
\providecommand \Eprint [0]{\href }%
\providecommand \doibase [0]{http://dx.doi.org/}%
\providecommand \selectlanguage [0]{\@gobble}%
\providecommand \bibinfo  [0]{\@secondoftwo}%
\providecommand \bibfield  [0]{\@secondoftwo}%
\providecommand \translation [1]{[#1]}%
\providecommand \BibitemOpen [0]{}%
\providecommand \bibitemStop [0]{}%
\providecommand \bibitemNoStop [0]{.\EOS\space}%
\providecommand \EOS [0]{\spacefactor3000\relax}%
\providecommand \BibitemShut  [1]{\csname bibitem#1\endcsname}%
\let\auto@bib@innerbib\@empty
\bibitem [{\citenamefont {Boccaletti}\ \emph {et~al.}(2006)\citenamefont
  {Boccaletti}, \citenamefont {Latora}, \citenamefont {Moreno}, \citenamefont
  {Chavez},\ and\ \citenamefont {Hwang}}]{boccaletti_complex_2006}%
  \BibitemOpen
  \bibfield  {author} {\bibinfo {author} {\bibfnamefont {S.}~\bibnamefont
  {Boccaletti}}, \bibinfo {author} {\bibfnamefont {V.}~\bibnamefont {Latora}},
  \bibinfo {author} {\bibfnamefont {Y.}~\bibnamefont {Moreno}}, \bibinfo
  {author} {\bibfnamefont {M.}~\bibnamefont {Chavez}}, \ and\ \bibinfo {author}
  {\bibfnamefont {D.-U.}\ \bibnamefont {Hwang}},\ }\href@noop {} {\bibfield
  {journal} {\bibinfo  {journal} {Phys. Rep.}\ }\textbf {\bibinfo {volume}
  {424}},\ \bibinfo {pages} {175} (\bibinfo {year} {2006})}\BibitemShut
  {NoStop}%
\bibitem [{\citenamefont {Watts}(2007)}]{watts_twenty-first_2007}%
  \BibitemOpen
  \bibfield  {author} {\bibinfo {author} {\bibfnamefont {D.~J.}\ \bibnamefont
  {Watts}},\ }\href {\doibase 10.1038/445489a} {\bibfield  {journal} {\bibinfo
  {journal} {Nature}\ }\textbf {\bibinfo {volume} {445}},\ \bibinfo {pages}
  {489} (\bibinfo {year} {2007})}\BibitemShut {NoStop}%
\bibitem [{\citenamefont {Lazer}\ \emph {et~al.}(2009)\citenamefont {Lazer},
  \citenamefont {Pentland}, \citenamefont {Adamic}, \citenamefont {Aral},
  \citenamefont {Barabási}, \citenamefont {Brewer}, \citenamefont
  {Christakis}, \citenamefont {Contractor}, \citenamefont {Fowler},
  \citenamefont {Gutmann}, \citenamefont {Jebara}, \citenamefont {King},
  \citenamefont {Macy}, \citenamefont {Roy},\ and\ \citenamefont
  {Alstyne}}]{lazer_computational_2009}%
  \BibitemOpen
  \bibfield  {author} {\bibinfo {author} {\bibfnamefont {D.}~\bibnamefont
  {Lazer}}, \bibinfo {author} {\bibfnamefont {A.}~\bibnamefont {Pentland}},
  \bibinfo {author} {\bibfnamefont {L.}~\bibnamefont {Adamic}}, \bibinfo
  {author} {\bibfnamefont {S.}~\bibnamefont {Aral}}, \bibinfo {author}
  {\bibfnamefont {A.-L.}\ \bibnamefont {Barabási}}, \bibinfo {author}
  {\bibfnamefont {D.}~\bibnamefont {Brewer}}, \bibinfo {author} {\bibfnamefont
  {N.}~\bibnamefont {Christakis}}, \bibinfo {author} {\bibfnamefont
  {N.}~\bibnamefont {Contractor}}, \bibinfo {author} {\bibfnamefont
  {J.}~\bibnamefont {Fowler}}, \bibinfo {author} {\bibfnamefont
  {M.}~\bibnamefont {Gutmann}}, \bibinfo {author} {\bibfnamefont
  {T.}~\bibnamefont {Jebara}}, \bibinfo {author} {\bibfnamefont
  {G.}~\bibnamefont {King}}, \bibinfo {author} {\bibfnamefont {M.}~\bibnamefont
  {Macy}}, \bibinfo {author} {\bibfnamefont {D.}~\bibnamefont {Roy}}, \ and\
  \bibinfo {author} {\bibfnamefont {M.~V.}\ \bibnamefont {Alstyne}},\ }\href
  {\doibase 10.1126/science.1167742} {\bibfield  {journal} {\bibinfo  {journal}
  {Science}\ }\textbf {\bibinfo {volume} {323}},\ \bibinfo {pages} {721}
  (\bibinfo {year} {2009})}\BibitemShut {NoStop}%
\bibitem [{\citenamefont {Guimerà}\ \emph {et~al.}(2010)\citenamefont
  {Guimerà}, \citenamefont {Stouffer}, \citenamefont {Sales-Pardo},
  \citenamefont {Leicht}, \citenamefont {Newman},\ and\ \citenamefont
  {Amaral}}]{guimera_origin_2010}%
  \BibitemOpen
  \bibfield  {author} {\bibinfo {author} {\bibfnamefont {R.}~\bibnamefont
  {Guimerà}}, \bibinfo {author} {\bibfnamefont {D.~B.}\ \bibnamefont
  {Stouffer}}, \bibinfo {author} {\bibfnamefont {M.}~\bibnamefont
  {Sales-Pardo}}, \bibinfo {author} {\bibfnamefont {E.~A.}\ \bibnamefont
  {Leicht}}, \bibinfo {author} {\bibfnamefont {M.~E.~J.}\ \bibnamefont
  {Newman}}, \ and\ \bibinfo {author} {\bibfnamefont {L.~A.~N.}\ \bibnamefont
  {Amaral}},\ }\href {\doibase 10.1890/09-1175.1} {\bibfield  {journal}
  {\bibinfo  {journal} {Ecology}\ }\textbf {\bibinfo {volume} {91}},\ \bibinfo
  {pages} {2941} (\bibinfo {year} {2010})}\BibitemShut {NoStop}%
\bibitem [{\citenamefont {Betzel}\ \emph {et~al.}(2013)\citenamefont {Betzel},
  \citenamefont {Griffa}, \citenamefont {Avena-Koenigsberger}, \citenamefont
  {Goñi}, \citenamefont {Thiran}, \citenamefont {Hagmann},\ and\ \citenamefont
  {Sporns}}]{betzel_multi-scale_2013}%
  \BibitemOpen
  \bibfield  {author} {\bibinfo {author} {\bibfnamefont {R.~F.}\ \bibnamefont
  {Betzel}}, \bibinfo {author} {\bibfnamefont {A.}~\bibnamefont {Griffa}},
  \bibinfo {author} {\bibfnamefont {A.}~\bibnamefont {Avena-Koenigsberger}},
  \bibinfo {author} {\bibfnamefont {J.}~\bibnamefont {Goñi}}, \bibinfo
  {author} {\bibfnamefont {J.-P.}\ \bibnamefont {Thiran}}, \bibinfo {author}
  {\bibfnamefont {P.}~\bibnamefont {Hagmann}}, \ and\ \bibinfo {author}
  {\bibfnamefont {O.}~\bibnamefont {Sporns}},\ }\href {\doibase
  10.1017/nws.2013.19} {\bibfield  {journal} {\bibinfo  {journal} {Netw. Sci.}\
  }\textbf {\bibinfo {volume} {1}},\ \bibinfo {pages} {353} (\bibinfo {year}
  {2013})}\BibitemShut {NoStop}%
\bibitem [{\citenamefont {Barabási}(2009)}]{barabasi_scale-free_2009}%
  \BibitemOpen
  \bibfield  {author} {\bibinfo {author} {\bibfnamefont {A.-L.}\ \bibnamefont
  {Barabási}},\ }\href {\doibase 10.1126/science.1173299} {\bibfield
  {journal} {\bibinfo  {journal} {Science}\ }\textbf {\bibinfo {volume}
  {325}},\ \bibinfo {pages} {412} (\bibinfo {year} {2009})}\BibitemShut
  {NoStop}%
\bibitem [{\citenamefont {Watts}\ and\ \citenamefont
  {Strogatz}(1998)}]{watts_collective_1998}%
  \BibitemOpen
  \bibfield  {author} {\bibinfo {author} {\bibfnamefont {D.~J.}\ \bibnamefont
  {Watts}}\ and\ \bibinfo {author} {\bibfnamefont {S.~H.}\ \bibnamefont
  {Strogatz}},\ }\href {\doibase 10.1038/30918} {\bibfield  {journal} {\bibinfo
   {journal} {Nature}\ }\textbf {\bibinfo {volume} {393}},\ \bibinfo {pages}
  {440} (\bibinfo {year} {1998})}\BibitemShut {NoStop}%
\bibitem [{\citenamefont {Ravasz}\ \emph {et~al.}(2002)\citenamefont {Ravasz},
  \citenamefont {Somera}, \citenamefont {Mongru}, \citenamefont {Oltvai},\ and\
  \citenamefont {Barabási}}]{ravasz_hierarchical_2002}%
  \BibitemOpen
  \bibfield  {author} {\bibinfo {author} {\bibfnamefont {E.}~\bibnamefont
  {Ravasz}}, \bibinfo {author} {\bibfnamefont {A.~L.}\ \bibnamefont {Somera}},
  \bibinfo {author} {\bibfnamefont {D.~A.}\ \bibnamefont {Mongru}}, \bibinfo
  {author} {\bibfnamefont {Z.~N.}\ \bibnamefont {Oltvai}}, \ and\ \bibinfo
  {author} {\bibfnamefont {A.-L.}\ \bibnamefont {Barabási}},\ }\href {\doibase
  10.1126/science.1073374} {\bibfield  {journal} {\bibinfo  {journal}
  {Science}\ }\textbf {\bibinfo {volume} {297}},\ \bibinfo {pages} {1551}
  (\bibinfo {year} {2002})}\BibitemShut {NoStop}%
\bibitem [{\citenamefont {Traud}, \citenamefont {Mucha},\ and\ \citenamefont
  {Porter}(2012)}]{traud_social_2012}%
  \BibitemOpen
  \bibfield  {author} {\bibinfo {author} {\bibfnamefont {A.~L.}\ \bibnamefont
  {Traud}}, \bibinfo {author} {\bibfnamefont {P.~J.}\ \bibnamefont {Mucha}}, \
  and\ \bibinfo {author} {\bibfnamefont {M.~A.}\ \bibnamefont {Porter}},\
  }\href {\doibase 10.1016/j.physa.2011.12.021} {\bibfield  {journal} {\bibinfo
   {journal} {Physica A}\ }\textbf {\bibinfo {volume} {391}},\ \bibinfo {pages}
  {4165} (\bibinfo {year} {2012})}\BibitemShut {NoStop}%
\bibitem [{\citenamefont {Fortunato}(2010)}]{fortunato_community_2010}%
  \BibitemOpen
  \bibfield  {author} {\bibinfo {author} {\bibfnamefont {S.}~\bibnamefont
  {Fortunato}},\ }\href {\doibase 10.1016/j.physrep.2009.11.002} {\bibfield
  {journal} {\bibinfo  {journal} {Phys. Rep.}\ }\textbf {\bibinfo {volume}
  {486}},\ \bibinfo {pages} {75} (\bibinfo {year} {2010})}\BibitemShut
  {NoStop}%
\bibitem [{\citenamefont {Newman}\ and\ \citenamefont
  {Girvan}(2004)}]{newman_finding_2004}%
  \BibitemOpen
  \bibfield  {author} {\bibinfo {author} {\bibfnamefont {M.~E.~J.}\
  \bibnamefont {Newman}}\ and\ \bibinfo {author} {\bibfnamefont
  {M.}~\bibnamefont {Girvan}},\ }\href {\doibase 10.1103/PhysRevE.69.026113}
  {\bibfield  {journal} {\bibinfo  {journal} {Phys. Rev. E}\ }\textbf {\bibinfo
  {volume} {69}},\ \bibinfo {pages} {026113} (\bibinfo {year}
  {2004})}\BibitemShut {NoStop}%
\bibitem [{\citenamefont {Duch}\ and\ \citenamefont
  {Arenas}(2005)}]{duch_community_2005}%
  \BibitemOpen
  \bibfield  {author} {\bibinfo {author} {\bibfnamefont {J.}~\bibnamefont
  {Duch}}\ and\ \bibinfo {author} {\bibfnamefont {A.}~\bibnamefont {Arenas}},\
  }\href {\doibase 10.1103/PhysRevE.72.027104} {\bibfield  {journal} {\bibinfo
  {journal} {Phys. Rev. E}\ }\textbf {\bibinfo {volume} {72}},\ \bibinfo
  {pages} {027104} (\bibinfo {year} {2005})}\BibitemShut {NoStop}%
\bibitem [{\citenamefont {Reichardt}\ and\ \citenamefont
  {Bornholdt}(2006)}]{reichardt_statistical_2006}%
  \BibitemOpen
  \bibfield  {author} {\bibinfo {author} {\bibfnamefont {J.}~\bibnamefont
  {Reichardt}}\ and\ \bibinfo {author} {\bibfnamefont {S.}~\bibnamefont
  {Bornholdt}},\ }\href {\doibase 10.1103/PhysRevE.74.016110} {\bibfield
  {journal} {\bibinfo  {journal} {Phys. Rev. E}\ }\textbf {\bibinfo {volume}
  {74}},\ \bibinfo {pages} {016110+} (\bibinfo {year} {2006})}\BibitemShut
  {NoStop}%
\bibitem [{\citenamefont {Guimerà}\ and\ \citenamefont
  {Nunes~Amaral}(2005)}]{guimera_functional_2005}%
  \BibitemOpen
  \bibfield  {author} {\bibinfo {author} {\bibfnamefont {R.}~\bibnamefont
  {Guimerà}}\ and\ \bibinfo {author} {\bibfnamefont {L.~A.}\ \bibnamefont
  {Nunes~Amaral}},\ }\href {\doibase 10.1038/nature03288} {\bibfield  {journal}
  {\bibinfo  {journal} {Nature}\ }\textbf {\bibinfo {volume} {433}},\ \bibinfo
  {pages} {895} (\bibinfo {year} {2005})}\BibitemShut {NoStop}%
\bibitem [{\citenamefont {Newman}(2006)}]{newman_finding_2006}%
  \BibitemOpen
  \bibfield  {author} {\bibinfo {author} {\bibfnamefont {M.~E.~J.}\
  \bibnamefont {Newman}},\ }\href {\doibase 10.1103/PhysRevE.74.036104}
  {\bibfield  {journal} {\bibinfo  {journal} {Phys. Rev. E}\ }\textbf {\bibinfo
  {volume} {74}},\ \bibinfo {pages} {036104+} (\bibinfo {year}
  {2006})}\BibitemShut {NoStop}%
\bibitem [{\citenamefont {Clauset}, \citenamefont {Newman},\ and\ \citenamefont
  {Moore}(2004)}]{clauset_finding_2004}%
  \BibitemOpen
  \bibfield  {author} {\bibinfo {author} {\bibfnamefont {A.}~\bibnamefont
  {Clauset}}, \bibinfo {author} {\bibfnamefont {M.~E.~J.}\ \bibnamefont
  {Newman}}, \ and\ \bibinfo {author} {\bibfnamefont {C.}~\bibnamefont
  {Moore}},\ }\href {\doibase 10.1103/PhysRevE.70.066111} {\bibfield  {journal}
  {\bibinfo  {journal} {Phys. Rev. E}\ }\textbf {\bibinfo {volume} {70}},\
  \bibinfo {pages} {066111} (\bibinfo {year} {2004})}\BibitemShut {NoStop}%
\bibitem [{\citenamefont {Blondel}\ \emph {et~al.}(2008)\citenamefont
  {Blondel}, \citenamefont {Guillaume}, \citenamefont {Lambiotte},\ and\
  \citenamefont {Lefebvre}}]{blondel_fast_2008}%
  \BibitemOpen
  \bibfield  {author} {\bibinfo {author} {\bibfnamefont {V.~D.}\ \bibnamefont
  {Blondel}}, \bibinfo {author} {\bibfnamefont {J.-L.}\ \bibnamefont
  {Guillaume}}, \bibinfo {author} {\bibfnamefont {R.}~\bibnamefont
  {Lambiotte}}, \ and\ \bibinfo {author} {\bibfnamefont {E.}~\bibnamefont
  {Lefebvre}},\ }\href {\doibase 10.1088/1742-5468/2008/10/P10008} {\bibfield
  {journal} {\bibinfo  {journal} {J. Stat. Mech.}\ }\textbf {\bibinfo {volume}
  {2008}},\ \bibinfo {pages} {P10008} (\bibinfo {year} {2008})}\BibitemShut
  {NoStop}%
\bibitem [{\citenamefont {Lancichinetti}\ and\ \citenamefont
  {Fortunato}(2009{\natexlab{a}})}]{lancichinetti_community_2009}%
  \BibitemOpen
  \bibfield  {author} {\bibinfo {author} {\bibfnamefont {A.}~\bibnamefont
  {Lancichinetti}}\ and\ \bibinfo {author} {\bibfnamefont {S.}~\bibnamefont
  {Fortunato}},\ }\href {\doibase 10.1103/PhysRevE.80.056117} {\bibfield
  {journal} {\bibinfo  {journal} {Phys. Rev. E}\ }\textbf {\bibinfo {volume}
  {80}},\ \bibinfo {pages} {056117} (\bibinfo {year}
  {2009}{\natexlab{a}})}\BibitemShut {NoStop}%
\bibitem [{\citenamefont {Traag}, \citenamefont {Van~Dooren},\ and\
  \citenamefont {Nesterov}(2011)}]{traag_narrow_2011}%
  \BibitemOpen
  \bibfield  {author} {\bibinfo {author} {\bibfnamefont {V.~A.}\ \bibnamefont
  {Traag}}, \bibinfo {author} {\bibfnamefont {P.}~\bibnamefont {Van~Dooren}}, \
  and\ \bibinfo {author} {\bibfnamefont {Y.}~\bibnamefont {Nesterov}},\ }\href
  {\doibase 10.1103/PhysRevE.84.016114} {\bibfield  {journal} {\bibinfo
  {journal} {Phys. Rev. E}\ }\textbf {\bibinfo {volume} {84}},\ \bibinfo
  {pages} {016114} (\bibinfo {year} {2011})}\BibitemShut {NoStop}%
\bibitem [{\citenamefont {Rosvall}\ and\ \citenamefont
  {Bergstrom}(2010)}]{rosvall_mapping_2010}%
  \BibitemOpen
  \bibfield  {author} {\bibinfo {author} {\bibfnamefont {M.}~\bibnamefont
  {Rosvall}}\ and\ \bibinfo {author} {\bibfnamefont {C.~T.}\ \bibnamefont
  {Bergstrom}},\ }\href {\doibase 10.1371/journal.pone.0008694} {\bibfield
  {journal} {\bibinfo  {journal} {PLoS ONE}\ }\textbf {\bibinfo {volume} {5}},\
  \bibinfo {pages} {e8694} (\bibinfo {year} {2010})}\BibitemShut {NoStop}%
\bibitem [{\citenamefont {Rosvall}\ and\ \citenamefont
  {Bergstrom}(2011)}]{rosvall_multilevel_2011}%
  \BibitemOpen
  \bibfield  {author} {\bibinfo {author} {\bibfnamefont {M.}~\bibnamefont
  {Rosvall}}\ and\ \bibinfo {author} {\bibfnamefont {C.~T.}\ \bibnamefont
  {Bergstrom}},\ }\href {\doibase 10.1371/journal.pone.0018209} {\bibfield
  {journal} {\bibinfo  {journal} {PLoS ONE}\ }\textbf {\bibinfo {volume} {6}},\
  \bibinfo {pages} {e18209} (\bibinfo {year} {2011})}\BibitemShut {NoStop}%
\bibitem [{\citenamefont {Ronhovde}\ and\ \citenamefont
  {Nussinov}(2010)}]{ronhovde_local_2010}%
  \BibitemOpen
  \bibfield  {author} {\bibinfo {author} {\bibfnamefont {P.}~\bibnamefont
  {Ronhovde}}\ and\ \bibinfo {author} {\bibfnamefont {Z.}~\bibnamefont
  {Nussinov}},\ }\href {\doibase 10.1103/PhysRevE.81.046114} {\bibfield
  {journal} {\bibinfo  {journal} {Phys. Rev. E}\ }\textbf {\bibinfo {volume}
  {81}},\ \bibinfo {pages} {046114} (\bibinfo {year} {2010})}\BibitemShut
  {NoStop}%
\bibitem [{\citenamefont {Evans}\ and\ \citenamefont
  {Lambiotte}(2009)}]{evans_line_2009}%
  \BibitemOpen
  \bibfield  {author} {\bibinfo {author} {\bibfnamefont {T.~S.}\ \bibnamefont
  {Evans}}\ and\ \bibinfo {author} {\bibfnamefont {R.}~\bibnamefont
  {Lambiotte}},\ }\href {\doibase 10.1103/PhysRevE.80.016105} {\bibfield
  {journal} {\bibinfo  {journal} {Phys. Rev. E}\ }\textbf {\bibinfo {volume}
  {80}},\ \bibinfo {pages} {016105} (\bibinfo {year} {2009})}\BibitemShut
  {NoStop}%
\bibitem [{\citenamefont {Lancichinetti}\ \emph {et~al.}(2011)\citenamefont
  {Lancichinetti}, \citenamefont {Radicchi}, \citenamefont {Ramasco},\ and\
  \citenamefont {Fortunato}}]{lancichinetti_finding_2011}%
  \BibitemOpen
  \bibfield  {author} {\bibinfo {author} {\bibfnamefont {A.}~\bibnamefont
  {Lancichinetti}}, \bibinfo {author} {\bibfnamefont {F.}~\bibnamefont
  {Radicchi}}, \bibinfo {author} {\bibfnamefont {J.~J.}\ \bibnamefont
  {Ramasco}}, \ and\ \bibinfo {author} {\bibfnamefont {S.}~\bibnamefont
  {Fortunato}},\ }\href {\doibase 10.1371/journal.pone.0018961} {\bibfield
  {journal} {\bibinfo  {journal} {PLoS ONE}\ }\textbf {\bibinfo {volume} {6}},\
  \bibinfo {pages} {e18961} (\bibinfo {year} {2011})}\BibitemShut {NoStop}%
\bibitem [{\citenamefont {Traag}, \citenamefont {Krings},\ and\ \citenamefont
  {Van~Dooren}(2013)}]{traag_significant_2013}%
  \BibitemOpen
  \bibfield  {author} {\bibinfo {author} {\bibfnamefont {V.~A.}\ \bibnamefont
  {Traag}}, \bibinfo {author} {\bibfnamefont {G.}~\bibnamefont {Krings}}, \
  and\ \bibinfo {author} {\bibfnamefont {P.}~\bibnamefont {Van~Dooren}},\
  }\href {\doibase 10.1038/srep02930} {\bibfield  {journal} {\bibinfo
  {journal} {Sci. Rep.}\ }\textbf {\bibinfo {volume} {3}},\ \bibinfo {pages}
  {2930} (\bibinfo {year} {2013})}\BibitemShut {NoStop}%
\bibitem [{\citenamefont {Traag}, \citenamefont {Aldecoa},\ and\ \citenamefont
  {Delvenne}(2015)}]{traag_detecting_2015}%
  \BibitemOpen
  \bibfield  {author} {\bibinfo {author} {\bibfnamefont {V.~A.}\ \bibnamefont
  {Traag}}, \bibinfo {author} {\bibfnamefont {R.}~\bibnamefont {Aldecoa}}, \
  and\ \bibinfo {author} {\bibfnamefont {J.-C.}\ \bibnamefont {Delvenne}},\
  }\href@noop {} {\bibfield  {journal} {\bibinfo  {journal} {arXiv:1503.00445
  [physics, stat]}\ } (\bibinfo {year} {2015})}\BibitemShut {NoStop}%
\bibitem [{\citenamefont {Barabási}\ and\ \citenamefont
  {Albert}(1999)}]{barabasi_emergence_1999}%
  \BibitemOpen
  \bibfield  {author} {\bibinfo {author} {\bibfnamefont {A.-L.}\ \bibnamefont
  {Barabási}}\ and\ \bibinfo {author} {\bibfnamefont {R.}~\bibnamefont
  {Albert}},\ }\href {\doibase 10.1126/science.286.5439.509} {\bibfield
  {journal} {\bibinfo  {journal} {Science}\ }\textbf {\bibinfo {volume}
  {286}},\ \bibinfo {pages} {509} (\bibinfo {year} {1999})}\BibitemShut
  {NoStop}%
\bibitem [{\citenamefont {Fortunato}\ and\ \citenamefont
  {Barthélemy}(2007)}]{fortunato_resolution_2007}%
  \BibitemOpen
  \bibfield  {author} {\bibinfo {author} {\bibfnamefont {S.}~\bibnamefont
  {Fortunato}}\ and\ \bibinfo {author} {\bibfnamefont {M.}~\bibnamefont
  {Barthélemy}},\ }\href {\doibase 10.1073/pnas.0605965104} {\bibfield
  {journal} {\bibinfo  {journal} {Proc. Natl. Acad. Sci. USA}\ }\textbf
  {\bibinfo {volume} {104}},\ \bibinfo {pages} {36} (\bibinfo {year}
  {2007})}\BibitemShut {NoStop}%
\bibitem [{\citenamefont {Aldecoa}\ and\ \citenamefont
  {Marín}(2013)}]{aldecoa_surprise_2013}%
  \BibitemOpen
  \bibfield  {author} {\bibinfo {author} {\bibfnamefont {R.}~\bibnamefont
  {Aldecoa}}\ and\ \bibinfo {author} {\bibfnamefont {I.}~\bibnamefont
  {Marín}},\ }\href {\doibase 10.1038/srep01060} {\bibfield  {journal}
  {\bibinfo  {journal} {Sci. Rep.}\ }\textbf {\bibinfo {volume} {3}},\ \bibinfo
  {pages} {1060} (\bibinfo {year} {2013})}\BibitemShut {NoStop}%
\bibitem [{\citenamefont {Lancichinetti}\ and\ \citenamefont
  {Fortunato}(2011)}]{Lancichinetti2011}%
  \BibitemOpen
  \bibfield  {author} {\bibinfo {author} {\bibfnamefont {A.}~\bibnamefont
  {Lancichinetti}}\ and\ \bibinfo {author} {\bibfnamefont {S.}~\bibnamefont
  {Fortunato}},\ }\href {\doibase 10.1103/PhysRevE.84.066122} {\bibfield
  {journal} {\bibinfo  {journal} {Phys. Rev. E}\ }\textbf {\bibinfo {volume}
  {84}},\ \bibinfo {pages} {066122} (\bibinfo {year} {2011})}\BibitemShut
  {NoStop}%
\bibitem [{\citenamefont {Lancichinetti}, \citenamefont {Fortunato},\ and\
  \citenamefont {Radicchi}(2008)}]{lancichinetti_benchmark_2008}%
  \BibitemOpen
  \bibfield  {author} {\bibinfo {author} {\bibfnamefont {A.}~\bibnamefont
  {Lancichinetti}}, \bibinfo {author} {\bibfnamefont {S.}~\bibnamefont
  {Fortunato}}, \ and\ \bibinfo {author} {\bibfnamefont {F.}~\bibnamefont
  {Radicchi}},\ }\href {\doibase 10.1103/PhysRevE.78.046110} {\bibfield
  {journal} {\bibinfo  {journal} {Phys. Rev. E}\ }\textbf {\bibinfo {volume}
  {78}},\ \bibinfo {pages} {046110} (\bibinfo {year} {2008})}\BibitemShut
  {NoStop}%
\bibitem [{\citenamefont {Moody}(2001)}]{Moody2001}%
  \BibitemOpen
  \bibfield  {author} {\bibinfo {author} {\bibfnamefont {J.}~\bibnamefont
  {Moody}},\ }\href {\doibase 10.1016/S0378-8733(01)00042-9} {\bibfield
  {journal} {\bibinfo  {journal} {Soc. Networks}\ }\textbf {\bibinfo {volume}
  {23}},\ \bibinfo {pages} {261} (\bibinfo {year} {2001})}\BibitemShut
  {NoStop}%
\bibitem [{\citenamefont {Cho}, \citenamefont {Myers},\ and\ \citenamefont
  {Leskovec}(2011)}]{Cho2011}%
  \BibitemOpen
  \bibfield  {author} {\bibinfo {author} {\bibfnamefont {E.}~\bibnamefont
  {Cho}}, \bibinfo {author} {\bibfnamefont {S.}~\bibnamefont {Myers}}, \ and\
  \bibinfo {author} {\bibfnamefont {J.}~\bibnamefont {Leskovec}},\ }in\ \href
  {\doibase 10.1145/2020408.2020579} {\emph {\bibinfo {booktitle} {Proceedings
  of the 17th ACM SIGKDD \ldots}}}\ (\bibinfo  {publisher} {ACM},\ \bibinfo
  {year} {2011})\ pp.\ \bibinfo {pages} {1082--1090}\BibitemShut {NoStop}%
\bibitem [{\citenamefont {Viswanath}\ \emph {et~al.}(2009)\citenamefont
  {Viswanath}, \citenamefont {Mislove}, \citenamefont {Cha},\ and\
  \citenamefont {Gummadi}}]{Viswanath2009}%
  \BibitemOpen
  \bibfield  {author} {\bibinfo {author} {\bibfnamefont {B.}~\bibnamefont
  {Viswanath}}, \bibinfo {author} {\bibfnamefont {A.}~\bibnamefont {Mislove}},
  \bibinfo {author} {\bibfnamefont {M.}~\bibnamefont {Cha}}, \ and\ \bibinfo
  {author} {\bibfnamefont {K.~P.}\ \bibnamefont {Gummadi}},\ }in\ \href
  {\doibase 10.1145/1592665.1592675} {\emph {\bibinfo {booktitle} {Proceedings
  of the 2nd ACM workshop on Online social networks - WOSN '09}}}\ (\bibinfo
  {publisher} {ACM Press},\ \bibinfo {address} {New York, New York, USA},\
  \bibinfo {year} {2009})\ p.~\bibinfo {pages} {37}\BibitemShut {NoStop}%
\bibitem [{\citenamefont {{Jure Leskovec, Jon
  Kleinberg}}(2006)}]{JureLeskovecJonKleinberg2006}%
  \BibitemOpen
  \bibfield  {author} {\bibinfo {author} {\bibfnamefont {C.~F.}\ \bibnamefont
  {{Jure Leskovec, Jon Kleinberg}}},\ }in\ \href@noop {} {\emph {\bibinfo
  {booktitle} {ACM Transactions on knowledge discovery from data}}}\ (\bibinfo
  {year} {2006})\ pp.\ \bibinfo {pages} {1--40}\BibitemShut {NoStop}%
\bibitem [{\citenamefont {Leskovec}\ \emph {et~al.}(2008)\citenamefont
  {Leskovec}, \citenamefont {Lang}, \citenamefont {Dasgupta},\ and\
  \citenamefont {Mahoney}}]{Leskovec2008}%
  \BibitemOpen
  \bibfield  {author} {\bibinfo {author} {\bibfnamefont {J.}~\bibnamefont
  {Leskovec}}, \bibinfo {author} {\bibfnamefont {K.~J.}\ \bibnamefont {Lang}},
  \bibinfo {author} {\bibfnamefont {A.}~\bibnamefont {Dasgupta}}, \ and\
  \bibinfo {author} {\bibfnamefont {M.~W.}\ \bibnamefont {Mahoney}},\ }in\
  \href {\doibase 10.1145/1367497.1367591} {\emph {\bibinfo {booktitle}
  {Proceeding of the 17th international conference on World Wide Web - WWW
  '08}}}\ (\bibinfo {year} {2008})\ p.\ \bibinfo {pages} {695},\ \Eprint
  {http://arxiv.org/abs/arXiv:0810.1355v1} {arXiv:arXiv:0810.1355v1}
  \BibitemShut {NoStop}%
\bibitem [{\citenamefont {Granovetter}(1973)}]{Granovetter1973}%
  \BibitemOpen
  \bibfield  {author} {\bibinfo {author} {\bibfnamefont {M.}~\bibnamefont
  {Granovetter}},\ }\href@noop {} {\bibfield  {journal} {\bibinfo  {journal}
  {Am. J. Sociol.}\ }\textbf {\bibinfo {volume} {78}},\ \bibinfo {pages} {1360}
  (\bibinfo {year} {1973})}\BibitemShut {NoStop}%
\bibitem [{\citenamefont {Onnela}\ \emph {et~al.}(2007)\citenamefont {Onnela},
  \citenamefont {Saram\"{a}ki}, \citenamefont {Hyv\"{o}nen}, \citenamefont
  {Szab\'{o}}, \citenamefont {Lazer}, \citenamefont {Kaski}, \citenamefont
  {Kert\'{e}sz},\ and\ \citenamefont {Barab\'{a}si}}]{Onnela2007a}%
  \BibitemOpen
  \bibfield  {author} {\bibinfo {author} {\bibfnamefont {J.}~\bibnamefont
  {Onnela}}, \bibinfo {author} {\bibfnamefont {J.}~\bibnamefont
  {Saram\"{a}ki}}, \bibinfo {author} {\bibfnamefont {J.}~\bibnamefont
  {Hyv\"{o}nen}}, \bibinfo {author} {\bibfnamefont {G.}~\bibnamefont
  {Szab\'{o}}}, \bibinfo {author} {\bibfnamefont {D.}~\bibnamefont {Lazer}},
  \bibinfo {author} {\bibfnamefont {K.}~\bibnamefont {Kaski}}, \bibinfo
  {author} {\bibfnamefont {J.}~\bibnamefont {Kert\'{e}sz}}, \ and\ \bibinfo
  {author} {\bibfnamefont {A.-L.}\ \bibnamefont {Barab\'{a}si}},\ }\href
  {\doibase 10.1073/pnas.0610245104} {\bibfield  {journal} {\bibinfo  {journal}
  {Proc. Natl. Acad. Sci. USA}\ }\textbf {\bibinfo {volume} {104}},\ \bibinfo
  {pages} {7332} (\bibinfo {year} {2007})}\BibitemShut {NoStop}%
\bibitem [{\citenamefont {Lancichinetti}\ and\ \citenamefont
  {Fortunato}(2009{\natexlab{b}})}]{Lancichinetti2009}%
  \BibitemOpen
  \bibfield  {author} {\bibinfo {author} {\bibfnamefont {A.}~\bibnamefont
  {Lancichinetti}}\ and\ \bibinfo {author} {\bibfnamefont {S.}~\bibnamefont
  {Fortunato}},\ }\href {\doibase 10.1103/PhysRevE.80.016118} {\bibfield
  {journal} {\bibinfo  {journal} {Physical Review E - Statistical, Nonlinear,
  and Soft Matter Physics}\ }\textbf {\bibinfo {volume} {80}},\ \bibinfo
  {pages} {1} (\bibinfo {year} {2009}{\natexlab{b}})},\ \Eprint
  {http://arxiv.org/abs/0904.3940} {arXiv:0904.3940} \BibitemShut {NoStop}%
\bibitem [{\citenamefont {Raghavan}, \citenamefont {Albert},\ and\
  \citenamefont {Kumara}(2007)}]{raghavan_near_2007}%
  \BibitemOpen
  \bibfield  {author} {\bibinfo {author} {\bibfnamefont {U.~N.}\ \bibnamefont
  {Raghavan}}, \bibinfo {author} {\bibfnamefont {R.}~\bibnamefont {Albert}}, \
  and\ \bibinfo {author} {\bibfnamefont {S.}~\bibnamefont {Kumara}},\ }\href
  {\doibase 10.1103/PhysRevE.76.036106} {\bibfield  {journal} {\bibinfo
  {journal} {Phys. Rev. E}\ }\textbf {\bibinfo {volume} {76}},\ \bibinfo
  {pages} {036106} (\bibinfo {year} {2007})}\BibitemShut {NoStop}%
\bibitem [{\citenamefont {Stanley}(2011)}]{stanley_enumerative_2011}%
  \BibitemOpen
  \bibfield  {author} {\bibinfo {author} {\bibfnamefont {R.~P.}\ \bibnamefont
  {Stanley}},\ }\href@noop {} {\emph {\bibinfo {title} {Enumerative
  {Combinatorics}: {Volume} 1}}},\ \bibinfo {edition} {2nd}\ ed.\ (\bibinfo
  {publisher} {Cambridge University Press},\ \bibinfo {address} {Cambridge,
  NY},\ \bibinfo {year} {2011})\BibitemShut {NoStop}%
\end{thebibliography}%

\balancecolsandclearpage

\appendix

\section{Complexity in a clique}

\label{sec:clique}

We here aim to determine the expected number of moves in the random neighbor algorithm. 
We assume it is always beneficial to move a node to a larger community.
In other words, whenever we select a random node $i$, and a random neighbor $j$, and the community $\sigma_j$ of the random neighbor is larger than the community $\sigma_i$ of node $i$, i.e. if $|V_{\sigma_j}| \geq |V_{\sigma_i}|$, we will move the node.

Let us denote by $f_k$ the number of communities that have size $k$
\begin{equation}
  f_k = |\{ c \mid |V_c| = k\}|.
\end{equation}
Then $g_k = kf_k$ denotes the number of nodes that belong to a community of size $k$.
Additionally, define $F_k = \sum_{i=k}^n f_i$ the number of communities that have size $k$ or larger.
Similarly, define $G_k = \sum_{i=k}^n g_i$ the number of nodes in communities that have size $k$ or larger.
Clearly $\sum_k g_k = n$ so that $G_1 = n$.
Also, $\sum_k f_k = r$ denotes the number of communities.
The probability to select a node from a community of size $k$ is then simply $\frac{g_k}{n}$. 
Let us denote by $X_{cd}$ the event of moving a node from a community of size $c$ to a community of size $d$.
Then the probability of $X_{cd}$ is 
\begin{equation}
  \Pr(X_{cd}) = \begin{cases}
    \frac{g_c}{n} \frac{g_d}{n} & \text{~if~} c < d \\
    \frac{g_c}{n} \frac{g_c - c}{n} & \text{~if~} c = d \\
    0 & \text{~if~} c > d \\
  \end{cases}
\end{equation}
The probability to move it from any community to any other is then $\sum_{cd} \Pr(X_{cd})$.
Alternatively, it is easy to see that the probability to move to any other community is $\frac{G_k - k}{n}$ (where we subtract $k$ to make sure it moves to another community, and not the same community).
So, overall, the probability we will move a node is then 
\begin{equation}
  \Pr(\text{move}) = \sum_{k=1}^n \frac{g_k}{n} \frac{G_k - k}{n}
\end{equation}
Similarly, the probability we will not move a node to another community (i.e. we remain stuck in the same partition) is then
\begin{equation}
  \Pr(\text{not move}) = \sum_{k=1}^n \frac{g_k}{n} \frac{n - G_k + k}{n}.
\end{equation}
We only reduce the number of communities if we move a node from a community of size $1$ of course.
Hence, the probability to reduce the number of communities by $1$ is then
\begin{equation}
  \Pr( \sum_i f_k = r - 1 ) = \frac{g_1}{n} \frac{G_k - 1}{n} 
  \label{equ:pr_reduce}
\end{equation}
Now it would be possible to construct a complete transition network from any of the partitions to other partitions.
However, this becomes quickly intractable, and rather difficult to solve.

Instead, we suggest to group partitions by the number of communities. 
Then, we divide the process into different phases.
The algorithm would be in phase $i$ whenever there are $n - i + 1$ communities.
In other words, in the first phase there are $n$ communities, and the next phase starts whenever one of these nodes is put in another community.
In the penultimate phase, there are only two communities.
Let us then denote by $t_i$ the number of moves during phase $i$, and by $t$ the total number of moves during the whole process.
We would like to examine $E(t)$, which we can write out as $\sum_i E(t_i)$ by linearity of expectation.

The number of ways to partition a set of $n$ nodes in $r$ sets can be denoted by $p_r(n)$, which is known as the partition function in number theory~\cite{stanley_enumerative_2011}.
This function obeys the recursive identity
\begin{equation}
  p_r(n) = p_r(n - r) + p_{r-1}(n-1),
\end{equation}
since there are $p_{r-1}(n-1)$ partitions with at least one community of size $1$ and $p_r(n-r)$ partitions that all have a community of size at least $2$ (since we put $r$ nodes in each one of the $r$ communities).
We would then like to know how many partitions there are that have $s$ communities of size $1$.
Let us first define $q_k(n)$ to denote the number of ways to partition $n$ into $k$ sets with at least one community of size $1$.
Secondly, we need its counterpart $u_k(n)$ which denotes the number of ways to partition $n$ into $k$ sets without any community of size $1$.
Obviously then $p_k(n) = q_k(n) + u_k(n)$.
We can then derive the recursion
\begin{align}
  q_k(n) &= \sum_{r = 1}^{k-1} u_{k-r}(n-r) \\
         &= \sum_{r = 1}^{k-1} p_{k-r}(n-r) - q_{k-r}(n - r)
\end{align}
The reasoning is as follows. 
If there are $r$ communities of size $1$, we should know how many partitions there are of $n - r$ nodes into $k - r$ communities without using communities of size $1$.
Here $q_1(1) = 1$.
More specifically, let us denote by $q_k(n, r)$ the number of partitions that have $r$ communities of size $1$ and in total $k$ communities, using $n$ nodes.
Then, obviously, $q_k(n) = \sum_{r=1}^{k-1} q_k(n, r)$.
Moreover, $q_k(n, r) = u_{k-r}(n-r) = p_{k-r}(n-r) - q_{k-r}(n - r)$.
The average probability to reduce the number of communities by $1$ is then
\begin{align}
  & \Pr\left(\sum f_s = k - 1 \mid \sum f_s = k\right) \\
  =& \sum_{r=1}^{k-1} \frac{q_k(n, r)}{p_k(n)} \frac{g_1}{n} \frac{G_1 - 1}{n} \\
  =& \sum_{r=1}^{k-1} \frac{q_k(n, r)}{p_k(n)} \frac{n - r}{n} \frac{n - 1}{n}
\end{align}
This expression is unfortunately not easy to evaluate analytically.
Moreover, it incorrectly assumes that each partition is equally likely \emph{a priori}, whereas we know that a more uneven distribution of community sizes is more likely due to the preferential attachment to the largest community.
However, the following upper bound is immediate 
\begin{multline}
  \Pr\left(\sum f_s = k - 1 \mid \sum f_s = k\right) \leq \\
    \frac{n - k + 1}{n} \frac{n - 1}{n}
  \label{equ:upper_bound}
\end{multline}

In general $E(t_i)$ can be calculated relatively straightforward as
\begin{equation}
  E(t_i) = \frac{1}{\Pr(\text{reduce community by~}1)} \\
\end{equation}
which using the bound in Eq.~(\ref{equ:upper_bound}) leads to
\begin{equation}
  E(t_i) \geq \frac{n}{n - k + 1} \frac{n}{n - 1}
\end{equation}
so that
\begin{align}
  E(t) &= \sum_i E(t_i) \geq \sum_{i=1}^n \frac{n}{n - i + 1}\frac{n}{n - 1} \\
       &\approx n \sum_{i=1}^n \frac{1}{i} \\
       &\approx \O( n \log n )
\end{align}

\end{document}